\documentclass{aastex}
\usepackage{spr-astr-addons}
\usepackage{url}

\def\kms{km~s$^{-1}$}
\def\cooz{$\rm{^{12}CO}~{\it J}$ = 1$-$0}
\def\tim{$\times$}
\def\hi{H{\footnotesize I}}
\def\hii{H{\footnotesize II}}
\def\sigmad{$\Sigma-D$}
\def\vlsr{$v_{\rm LSR}$}

\begin{document}

\title{SRAO CO Observation of 11 Supernova Remnants \\ in $l$ = 70$^\circ$ to 190$^\circ$}

\shorttitle{SRAO CO Observation of 11 Supernova Remnants in $l$ = 70$^\circ$ to 190$^\circ$}
\shortauthors{Jeong et al.}

\author{Il-Gyo Jeong\altaffilmark{1}}
\affil{Astronomy Program, Department of Physics and Astronomy,
Seoul National University, Seoul 151-742, Republic of Korea}
\email{igjeong@astro.snu.ac.kr}

\and
\author{Do-Young Byun\altaffilmark{2,3}}
\affil{Korea Astronomy \& Space Science Institute,
Daejeon 305-348, Republic of Korea}
\email{bdy@kasi.re.kr}
\and

%\author{Do-Young Byun\altaffilmark{2,3}}
\affil{Yonsei University Observatory, Yonsei University, Seongsan-ro 262,
Seodaemun, Seoul 120-749, Republic of Korea}
\and

\author{Bon-Chul Koo\altaffilmark{1}}
\affil{Astronomy Program, Department of Physics and Astronomy,
Seoul National University, Seoul 151-742, Republic of Korea}
\email{koo@astro.snu.ac.kr}
\and

\author{Jea-Joon Lee, Jung-Won Lee, Hyunwoo Kang\altaffilmark{2}}
\affil{Korea Astronomy \& Space Science Institute,
Daejeon 305-348, Republic of Korea}
\email{leejjoon@kasi.re.kr, jwl@kasi.re.kr, orionkhw@kasi.re.kr}

%\altaffiltext{1}{Astronomy Program, Department of Physics and Astronomy,
%Seoul National University, Seoul 151-742, Republic of Korea}

%\altaffiltext{2}{Korea Astronomy \& Space Science Institute,
%Daejeon 305-348, Republic of Korea}

%\altaffiltext{3}{Yonsei University Observatory, Yonsei University, Seongsan-ro 262,
%Seodaemun, Seoul 120-749, Republic of Korea}

%--------------------------------Abstract start------------------------------------
\abstract{We present the results of $\rm{^{12}CO}~{\it J}$ = 1$-$0 line
observations of eleven Galactic supernova remnants (SNRs)
obtained using the Seoul Radio Astronomy Observatory (SRAO) 6-m radio telescope.
The observation was made as a part of the SRAO CO survey of SNRs
between $l$ = 70$^\circ$ and 190$^\circ$, which is intended
to identify SNRs interacting with molecular clouds.
The mapping areas for the individual SNRs are determined to cover their full extent
in the radio continuum. We used half-beam grid spacing (60$''$) for 9 SNRs
and full-beam grid spacing (120$''$) for the rest.
We detected CO emission towards most of the remnants.
In six SNRs, molecular clouds showed a good spatial relation with their radio
morphology, although no direct evidence for the interaction was detected.
Two SNRs are particularly interesting:
G85.4+0.7, where there is a filamentary molecular cloud
along the radio shell, and
3C434.1, where a large molecular cloud appears to block
the western half of the remnant. We briefly summarize the results
obtained for
individual SNRs.}

%--------------------------------Abstract end------------------------------------

%---------------------------------------------Introduction start-----------------------------------
\section{Introduction}\label{s:intro}
Supernova (SN) explosions are one of the most energetic events that can occur in a galaxy.
SN explosions strongly modify the environment, and at the same time,
the evolution of a supernova remnant (SNR)
is governed primarily by the environment itself.
As stars are formed due to the gravitational collapse of molecular clouds,
there is a large possibility that a SN explodes adjacent to its parental
molecular cloud (MC). Thus, an environment with molecular clouds may not be unusual
during a SNR evolution.

There are indeed SNRs that show direct evidence of SNR
interaction with molecular clouds
\citep[e.g., see][]{koo03a,jiang10}.
IC443 is a prototypical object from which
high-velocity, broad emission lines from
shocked atomic and molecular gases have been detected
\citep[][and references therein]{denoyer79,snell05,lee08,lee12}.
Shock-excited infrared emission lines from H$_2$ and
other molecules have been detected as well.
On the other hand, it is
proposed that centrally enhanced thermal X-ray emissions or the spatial coincidence
of molecular clouds can serve as an indirect indication of SNR-MC interaction.
SNR 3C391 belongs to this category, having break-out morphology in radio
emission \citep{reach99,chen04}.

%---------------table 1-------------------------------------------------------------------------

\begin{table*}
\normalsize
\caption{List of Supernova Remnants between $l$=70$^\circ$ and 190$^\circ$\label{tbl-1}}
\begin{tabular}{@{}ccccccc@{}}
\tableline
~~~~~Galactic~~~   &~~~Name(s)~~&~~~$\alpha$(J2000)~~&~~~$\delta$(2000)~~ &~~~Size~~    &~~~Type$^{a}$~~&~~~CO$^{b}$~~~~~\\
~~~~~Coordinates~~~&~~~    ~~   &~~~(h m s)        ~~&~~~($^\circ$~~$'$)~~&~~~(arcmin)~~&~~~~~          &~~~Observation~~~~~\\
\tableline
73.9 $+$0.9   &              &20 14 15           &+36 12            &22?          &S?         &This paper\\
74.0 $-$8.5   &Cygnus Loop   &20 51 00           &+30 40            &230\tim160   &S          &---\\
74.9 $+$1.2   &CTB87         &20 16 02           &+37 12            &8\tim6       &F          &4 \\
76.9 $+$1.0   &              &20 22 20           &+38 43            &12\tim9      &?          &This paper\\
78.2 $+$2.1   &$\gamma$~Cygni SNR &20 20 50      &+40 26            &60           &S          &1, 5\\
\\
82.2 $+$5.3   &W63           &20 19 00           &+45 30            &95\tim65     &S          &1\\
84.2 $-$0.8   &              &20 53 20           &+43 27            &20\tim16     &S          &This paper, 6\\
84.9 $+$0.5$^{c}$ &              &20 50 30           &+44 53            &6            &S          &---\\
85.4 $+$0.7   &              &20 50 40           &+45 22            &24           &S          &This paper\\
85.9 $-$0.6   &              &20 58 40           &+44 53            &24           &S          &This paper\\
\\
89.0 $+$4.7   &HB21          &20 45 00           &+50 35            &120\tim90    &S          &1, 2\\
93.3 $+$6.9   &DA530         &20 52 25           &+55 21            &27\tim20     &S          &This paper\\
93.7 $-$0.2   &CTB104A       &21 29 20           &+50 50            &80           &S          &3\\
94.0 $+$1.0   &3C434.1       &21 24 50           &+51 53            &30\tim25     &S          &This paper\\
106.3 $+$2.7  &              &22 27 30           &+60 50            &60\tim24     &?          &3, 7\\
\\
109.1 $-$1.0  &CTB109        &23 01 35           &+58 53            &28           &S          &3, 8\\
111.7 $-$2.1  &Cassiopeia A  &23 23 26           &+58 48            &5            &S          &3\\
114.3 $+$0.3  &              &23 37 00           &+61 55            &90\tim55     &S          &3\\
116.5 $+$1.1  &              &23 53 40           &+63 15            &80\tim60     &S          &3\\
116.9 $+$0.2  &CTB 1         &23 59 10           &+62 26            &34           &S          &1, 3\\
\\
119.5 $+$10.2 &CTA 1         &00 06 40           &+72 45            &90?          &S          &1\\
120.1 $+$1.4  &Tycho         &00 25 18           &+64 09            &8            &S          &3\\
126.2 $+$1.6  &              &01 22 00           &+64 15            &70           &S?         &3\\
127.1 $+$0.5  &R5            &01 28 20           &+63 10            &45           &S          &3\\
130.7 $+$3.1  &3C58          &02 05 41           &+64 49            &9\tim5       &F          &3\\
\\
132.7 $+$1.3  &HB3           &02 17 40           &+62 45            &80           &S          &1, 3\\
156.2 $+$5.7  &              &04 58 40           &+51 50            &110          &S          &1\\
160.9 $+$2.6  &HB9           &05 01 00           &+46 40            &140\tim120   &S          &1\\
166.0 $+$4.3  &VRO 42.05.01  &05 26 30           &+42 56            &55\tim35     &S          &1\\
166.2 $+$2.5$^{c}$  &OA 184        &05 19 00           &+41 55            &90\tim70     &S          &This paper\\
\\
179.0 $+$2.6  &              &05 53 40           &+31 05            &70           &S?         &This paper\\
180.0 $-$1.7  &S147          &05 39 00           &+27 50            &180          &S          &This paper\\
182.4 $+$4.3  &              &06 08 10           &+29 00            &50           &S          &This paper\\
184.6 $-$5.8  &Crab Nebula   &05 34 31           &+22 01            &7\tim5       &F          &---\\
189.1 $+$3.0  &IC443         &06 17 00           &+22 34            &45           &C          &3, 9\\
\tableline
\end{tabular}
\tablenotetext{a}{Type of the SNR (S: Shell, F: Filled-Centre, C: Composite) \citep{green04}}
\tablenotetext{b}{References are: (1) \cite{byun04}, (2) \cite{byun06}, (3) FCRAO Survey \citep{taylor03},
                 (4) \cite{kothes03}, (5) \cite{higgs83}, (6) \cite{feldt93}, (7) \cite{kothes01b},
                 (8) \cite{sasaki06}, (9) \cite{snell05}.}
\tablenotetext{c}{G84.9+0.5 and G166.2+2.5 (OA184) have been identified as \hii\ regions (Foster et al. 2006, 2007) and removed in the Green's 2009 catalog.}
\tablecomments{The parameters of the SNRs are from the Green's 2004 catalog \citep{green04}.}

\end{table*} 
%---------------table 1-------------------------------------------------------------------------

%---------------------------------------------table start----------------------------------
\begin{table}[t]
\small
%\begin{center}
\caption{Mapping Parameters of SNRs\label{tbl-2}}
\begin{tabular}{@{}ccccc@{}}
\tableline
Galactic   & Name(s)&Size     &Observed    &Grid \\
Coordinates&        &         &area        &spacing    \\
           &        &(arcmin) &(arcmin)    &(arcsec)            \\
\tableline
73.9 $+$0.9 &          &22?        &53\tim41             &60\\
76.9 $+$1.0 &          &12\tim9    &17\tim18             &60\\
84.2 $-$0.8 &          &20\tim16   &33\tim27             &60\\
85.4 $+$0.7 &          &24         &51\tim61             &60\\
85.9 $-$0.6 &          &24         &37\tim40             &60\\
93.3 $+$6.9 &DA530     &27\tim20   &37\tim36             &60\\
94.0 $+$1.0 &3C434.1   &30\tim25   &42\tim39             &60\\
166.2 $+$2.5&OA184     &90\tim70   &97\tim94             &60\\
179.0 $+$2.6&          &70         &108\tim130           &120\\
180.0 $-$1.7&S147      &180        &252\tim274           &120\\
182.4 $+$4.3&          &50         &63\tim61             &60\\
\tableline
\end{tabular}

%\end{center}
\end{table}

%---------------------------------------------table end----------------------------------

SNR-MC interactions have several
important astrophysical implications.
Strong SNR shocks interacting with molecular clouds may produce
high-energy gamma-ray emission due to
enhanced pion decay emission from the collision of
cosmic ray protons accompanied with efficient kinematic
energy conversion in a dense molecular environment.
The gamma-ray emission detected towards
the SNRs CTB 37B, IC 443, and W51C indeed show an extended spatial distribution
coincident with dense molecular clouds
\citep{aharonian08,acciari09,abdo09}. This could be the first
direct evidence for the production of cosmic ray protons by SNR shocks.
The shocked, dense molecular gas could provide a physical condition
for maser emission, e.g., 1720-MHz
maser line ($^2\Pi_{3/2}$, $J={3\over2}$, $F=2\rightarrow 1$) of the
OH molecule.
From SNRs W28, W44, and IC443,
OH maser emissions were detected behind the leading edge of the shock
\citep{claussen97}, and
according to theoretical calculations
\citep{elitzur76,lockett99,wardle99},
the masers arise only in slow, non-dissociative $C$-shocks
with a large OH column density, which can be met only when the shock is
observed tangentially
\citep[see the review by][]{wardle02}.

About 270 SNRs are known in our Galaxy \citep{green09},
and presently, about 20\% -- 30\% of them are known to be interacting with
molecular clouds \citep{jiang10}.
However, insufficient effort has been made to conduct systematic searches.
\cite{Huang86} studied a molecular environment of galactic
SNRs in $l$ = $70^\circ$ to $210^\circ$. However, due to the insufficient angular
resolution ($\sim$8$'$.7) of the telescope,
the results were only useful for studying the
distributions of large molecular cloud complexes near SNRs.
More recently, large-scale CO surveys of the galactic plane have been conducted
such as the Outer Galaxy Survey ($l = $ 102$^\circ$.5 -- 141$^\circ$.5)
or the Galactic Ring Survey
using the 14-m telescope of the Five College Radio Astronomy Observatory
with an FHWM of 50$'$$'$ \citep{taylor03,Jackson06}.
There have been studies on individual SNRs in this survey area; however,
no systematic studies have been conducted.

On the basis of the above background, we decided to carry out a systematic CO study
of SNRs using the Seoul Radio Astronomy Observatory (SRAO) 6-m telescope.
We limited the targets to the SNRs in $l$ = 70$^\circ$ to 190$^\circ$.
The targets in the inner galaxy were excluded because of the
ambiguity in determining the association.
We observed most of the SNRs listed the Green's 2004
catalog \citep{green04}.
Some results of the survey were reported previously by
\cite{byun04} and \cite{byun06}.
They observed 9 SNRs having bright X-ray emissions and large angular sizes.
They detected broad-line CO emissions from two SNRs (HB 21 and HB 3) and
also found a morphological correlation between the CO and radio continuum distribution for
two other SNRs ($\gamma$-Cygni and HB 9). For the remaining five SNRs,
no evidence of SNR-MC interaction was observed.
In this paper, we present the results obtained for eleven other SNRs.

This paper is organized as follows. In $\S$~\ref{s:obs},
the observation details and target selection are explained.
In $\S$~\ref{s:result}, the results for individual SNRs are presented.
Section~\ref{s:summary} summarizes the main results of this paper.
%---------------------------------------------Introduction end-----------------------------------

%---------------------------------------------Observation start----------------------------------
\section{Observations}\label{s:obs}

$^{12}$CO J$=$1$-$0 (115.271204 GHz) observations were carried out from October 2003 to
May 2005 using SRAO 6-m telescope.
The telescope has an FWHM beam size of 120$'$$'$ and a
main beam efficiency of 0.70 at 115 GHz \citep{koo03b}.
We used a 100 GHz SIS mixer receiver with a single-side band filter \citep{lee02}
and a 1024-channel auto-correlator with 50 MHz bandwidth, corresponding to a velocity coverage of 130~\kms~\citep{choi03}.
The system temperature ranged from 500 to 800 K
depending on the elevation and weather conditions.
Typical $rms$ noise level on T$_{mb}$ scale was $\sim$0.3 K at a 1 \kms\
velocity resolution.
In order to check the system performance, we regularly observed the bright standard source near
the target at one or two hour intervals.
Pointing accuracy was better than 20$'$$'$.
The data were reduced by using the CLASS\footnote{\url{http://www.iram.fr/IRAMFR/GILDAS}}
software in the Gildas package developed by IRAM.

%---------------fig 1---------------------------------------------------------------------------
\begin{figure*}[!t]
\hspace{1.cm}
\includegraphics[scale=0.9]{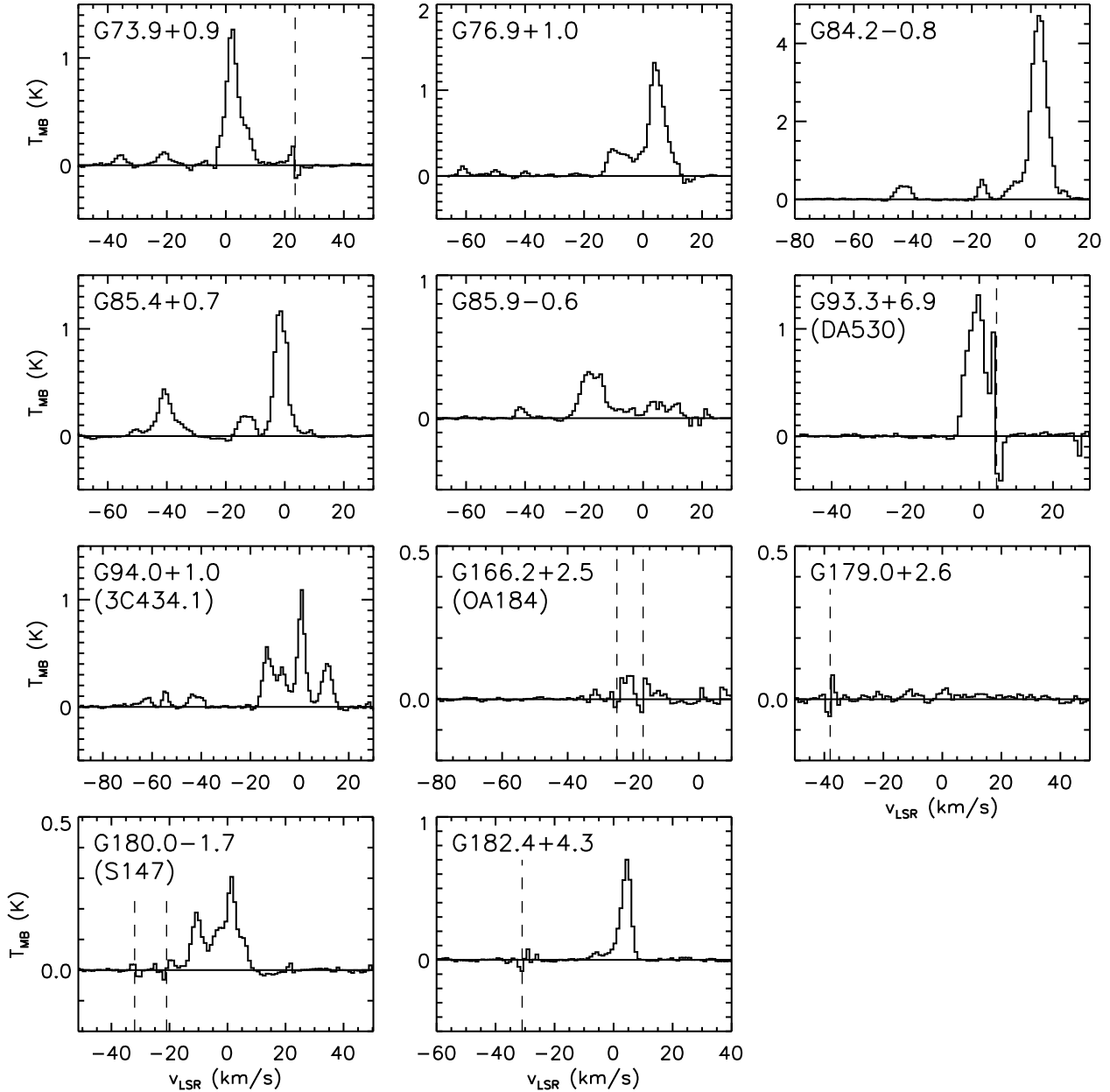}
\caption{Average $\rm{^{12}CO}~{\it J}$ = 1$-$0~spectra of 11 supernova remnants.
         The dashed lines mark the telluric CO line positions.
\label{fig:aver_spec}}
\end{figure*}
%---------------fig 1---------------------------------------------------------------------------

The targets were selected as follows.
First, the Galactic longitudes of the SNRs are limited to $l$ = 70$^\circ$ to 190$^\circ$.
Toward the inner Galactic region, many clouds are superposed along
the line-of-sight, and hence, it is difficult to confirm the association
of SNRs with molecular clouds.
On the other hand, the SNRs beyond 190$^\circ$ are at low declinations due to which
we have a rather limited observing time for them.
According to the Green's SNR catalog \citep{green04}, there are 35 SNRs in this longitude range,
and they are listed in Table~\ref{tbl-1}\footnote{In the Green's 2009 catalog \citep{green09}, the number has
increased to 37.}.
Second, the angular sizes of the SNRs are limited
to $\ge$10$'$ and $\le$180$'$, considering the
beam size of the SRAO 6-m telescope. If the angular size is excessively small,
the morphological comparison between the SNR and the molecular cloud
becomes difficult, whereas if it is excessively large, the observation
becomes highly time-consuming.
Third, the SNRs in the area considered by the
Five College Radio Astronomy Observatory Outer Galaxy Survey (FCRAO OGS) are excluded
because the survey was performed with
an adequate angular resolution ($60$$'$$'$) and sensitivity.
The FCRAO OGS survey
covered the region 102$^\circ$.5$<$$l$$<$141$^\circ$.5 and $-$3$^\circ$$<$$b$$<$5$^\circ$.4
\citep{taylor03}.
Finally, we excluded the SNRs observed by previous SRAO \cooz~observations
(Byun 2004, Byun et al. 2006). In the last column of Table~\ref{tbl-1}, we list the
references on the CO observations of individual SNRs.

The mapping areas for the individual SNRs are determined to cover their
full extents in the radio continuum. Nine SNRs were observed with half-beam grid spacing (60$'$$'$), whereas the two remaining SNRs were observed with full-beam grid spacing (120$'$$'$) via the position switching mode.
Table~\ref{tbl-2} lists the observed areas and the grid spacings for the target SNRs.
The total number of spectra obtained in this study is $\sim$45000.
%---------------------------------------------Observation end----------------------------------

%---------------------------------------------Result start----------------------------------------
\section{Results}\label{s:result}
Figure~\ref{fig:aver_spec} shows the average spectra toward 11 SNRs. Strong CO emission lines are
detected toward all SNRs other than G166.2+2.5 and G179.0+2.6.
In the following subsections, we summarize the results obtained for individual SNRs.

%----------------------------------fig 2-----------------------------------------------
\begin{figure}[t!]
\hspace{0.3cm}
\includegraphics[scale=0.82]{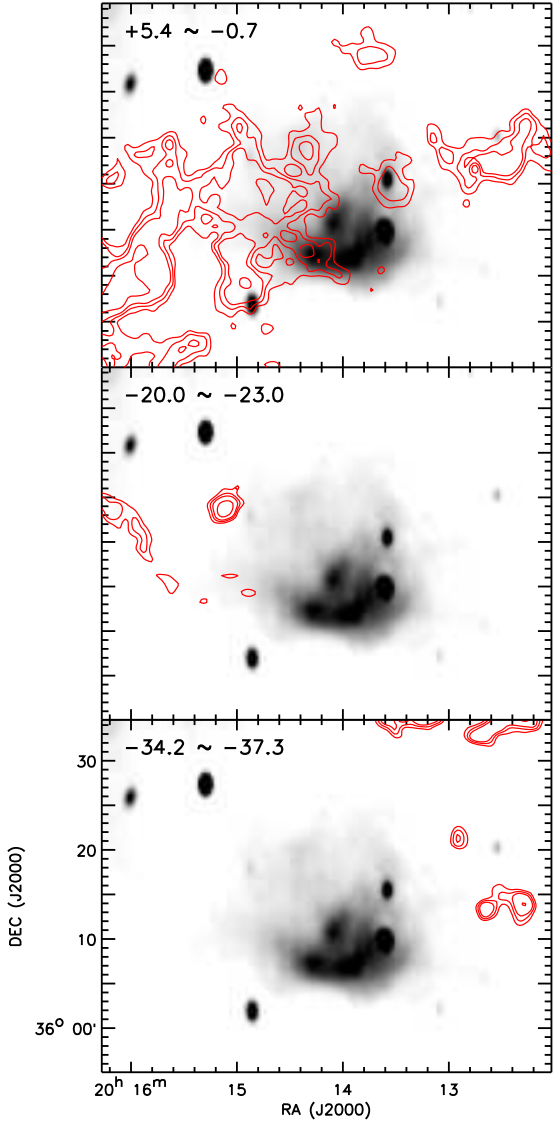}
\caption{$\rm{^{12}CO}~{\it J}$ = 1$-$0~average intensity maps of G73.9+0.9 (contour levels: 1.2, 1.6, 2, 3 K).
The gray-scale image is the Canadian Galactic Plane Survey (CGPS) 1420 MHz radio continuum image (scale range: 11$\sim$21 K) \citep{taylor03}.
\label{fig:g73}}
\end{figure}
%---------------------------------fig 2------------------------------------------------

%---------------------------------fig 3------------------------------------------------
\begin{figure}[t!]
\hspace{-0.27cm}
\includegraphics[scale=0.478]{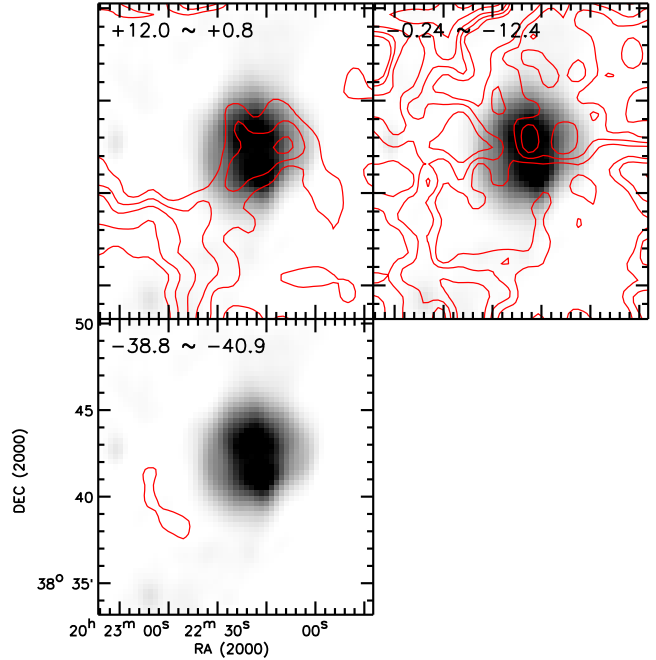}
\caption{$\rm{^{12}CO}~{\it J}$ = 1$-$0~average intensity maps of G76.9+1.0 (contour levels: 0.5, 0.8, 1.2, 2, 3 K).
The gray-scale image is the CGPS 1420 MHz radio continuum image (scale range: 11$\sim$21 K).
\label{fig:g76}}
\end{figure}
%---------------------------------fig 3------------------------------------------------

\subsection{G73.9+0.9}
G73.9+0.9 is located in the complex Cygnus area in the sky.
This source has a faint and broad, spur-like structure in radio and
is classified as an SNR on the basis of its polarization and
spectral index \citep{reich86}.
According to a recent radio continuum study,
the spectral index of this target is 0.23 and
there could be a pulsar wind nebula in the central region \citep{kothes06}.
In \hi, \cite{pineault96} suggested that there are no obvious
\hi\ features associated with the remnant, although large-scale \hi\ features
at $-$45~\kms\ indicate a complementary morphology.
\cite{case98} suggested a distance of 6.6 kpc, based on the \sigmad~relation.

According to our survey, CO molecular clouds appeared at around $+$2, $-$21, and $-$36~\kms~(Fig.~\ref{fig:aver_spec}).
The $+$2~\kms~component was the most prominent (Fig.~\ref{fig:g73}).
At this velocity, there is a large and filamentary cloud crossing the northern
part of the SNR along east-west direction.
The southern part of the SNR is bright in radio continuum,
whereas the northern part, where the CO cloud overlaps, is faint.
However, there is no evidence of the interaction between the molecular cloud
and the SNR.
Although we detected a few clumpy clouds at other velocities,
they showed no correlation with the SNR.

%---------------------------------fig 4------------------------------------------------
\begin{figure}[t!]
\hspace{0.cm}
\includegraphics[scale=0.82]{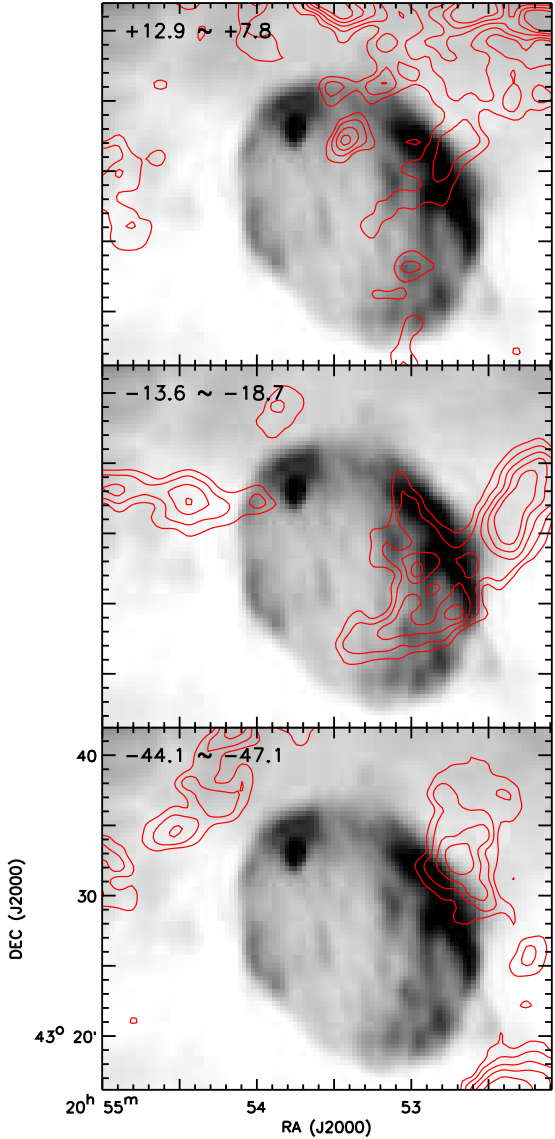}
\caption{$\rm{^{12}CO}~{\it J}$ = 1$-$0~average intensity maps of G84.2$-$0.8
(contour levels: 0.5, 0.8, 1.2 ,1.6 ,2 K).
The gray-scale image is the CGPS 1420 MHz radio continuum image (scale range: 11$\sim$24 K).
\label{fig:g84}}
\end{figure}
%---------------------------------fig 4------------------------------------------------

\subsection{G76.9+1.0}
This source is the smallest ($\sim$10$'$) among our targets.
It is a filled-center SNR with a diffuse extended envelope
in the radio continuum, and it could be
a pulsar wind nebula \citep{landecker93,kothes06}.
\cite{landecker93} suggested a distance greater than 7 kpc based on its
large rotation measure.
Based on the \sigmad~relation, \cite{case98} estimated the distance to be about 4.7 kpc.

We find that the most prominent CO component that appears at $\sim$+5~\kms\ is
distributed near the center and the southeast region of the remnant
(Figs.~\ref{fig:aver_spec} \&~\ref{fig:g76}).
However, this velocity component shows no spatial correlation with the radio continuum.
Several CO clouds are also detected around the SNR at a \vlsr\ from $-$12 to 0 \kms\
and near the south-east boundary at $-$40 \kms.
However, there is no obvious morphological correlation between the radio
and CO distribution.

\subsection{G84.2$-$0.8}
G84.2$-$0.8 is shell-type remnant having filamentary structures parallel to
the major axis with no polarized emission \citep{matthews80,kothes06}.
\cite{Huang86} showed that this remnant
lied at the south of a large molecular cloud complex
at a \vlsr\ between $-$44 and $-$33 \kms.
They estimated the kinematic distance to be 7.2 kpc and the total mass to be
$8.5 \times 10^5$ M$_\odot$ for the cloud. \cite{feldt93} performed \cooz~observation
with the KOSMA 3-m telescope (FWHM = $3.'8$)
in addition to \hi\ observations using the DRAO Synthesis Telescope (FWHM $\approx$ 1$'\times1.'5$).
They found that the CO molecular cloud at $-$17 \kms,
corresponding to a kinematic distance of $\sim$4 kpc is spatially coincident with the SNR.
No OH maser emission was detected toward the remnant \citep{frail96}.

We detected CO emission at four velocities: $\sim$ +11, +2.7, $-$17, and $-$46 \kms~(Fig.~\ref{fig:aver_spec}).
The $-17$ \kms\ cloud detected by \cite{feldt93} is more clearly observed in our data
(Fig.~\ref{fig:g84}). There are two clouds: a large one in the west and
a small one in the east.
The large cloud in the west has a semi-circular shape and is localized inside the remnant.
Its northwest boundary matches well with the boundary of the remnant with no asymmetric line emission.
The radio continuum is enhanced in the region where a molecular cloud is located.
These features suggest a possible
interaction between the molecular cloud and the remnant.
In order to confirm the interaction, we need to obtain more observations using higher transition CO lines.
If this cloud were interacting with the SNR,
the distance to the SNR would be 4.9 kpc according to
the rotation curve of Brand \& Blitz (1993).
The velocity component of $-$46 \kms\ appears to be aligned along the southwest
and the west. However, it does not show any correlated features with the remnant.

%---------------------------------fig 5------------------------------------------------
\begin{figure}[t!]
\hspace{-0.2cm}
\includegraphics[scale=0.51]{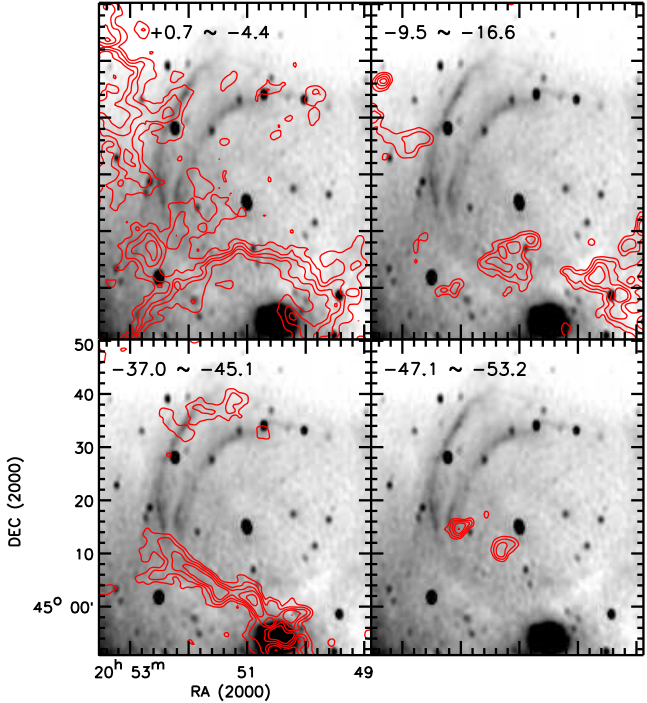}
\caption{$\rm{^{12}CO}~{\it J}$ = 1$-$0~average intensity maps of G85.4$+$0.7
(contour levels: 0.6, 1.0, 1.5, 2, 3 K).
The gray-scale image is the CGPS 1420 MHz radio continuum image (scale range: 8.5$\sim$11.5 K).
\label{fig:g85_4}}
\end{figure}
%---------------------------------fig 5------------------------------------------------

\subsection{G85.4+0.7}

G85.4+0.7 is shell-type SNR with two separated faint shells, discovered recently
from the Canadian Galactic Plane Survey (CGPS) \citep{kothes01a}.
Kothes et al. (2001a) detected a large \hi\ loop structure surrounding the SNR
at a \vlsr\ of $-$12 \kms. They interpreted the structure as a stellar-wind shell
generated by the progenitor of the SNR, and proposed a distance of 3.8 kpc to the SNR
corresponding to the velocity of the \hi\ loop.

We detected CO emissions at $\sim$ $-$1, $-$12, $-$41, and $-$50 \kms~
(Fig.~\ref{fig:aver_spec}). The molecular cloud at $-41$ \kms~
shows a considerably interesting feature (Fig.~\ref{fig:g85_4}).
This molecular cloud is in filamentary shape
and aligned along the south-east boundary of the SNR.
The spatial relation between the cloud and the SNR suggests that 
their association is very likely. 
The cloud extends to the bright, compact radio source below
the remnant, which is the \hii\ region G84.9$+$0.5 \citep{foster07}.
The velocity of the \hii\ region 
from radio recombination line observation is $v_{\rm LSR}=-39.3$~\kms, and 
\citet{foster07} proposed a kinematic distance of 
4.9 kpc considering the noncircular motions near the Perseus spiral arm.
The velocity of the CO cloud ($-$41 \kms) agrees with that of the radio recombination line,
and the cloud surrounds the \hii\ region, which
indicates that they are physically associated.
The CO emission lines in this area show asymmetric profiles that are
probably related to the outflows of young stellar objects.
If the $-41$~\kms\ molecular cloud were physically associated with both 
the \hii\ region and the SNR, the distance to the SNR would be 
4.9 kpc \citep{foster07} or 7.2 kpc if we adopt the rotation curve of \citet{brand93}.
And it will be difficult to associate the \hi\ loop structure detected by 
\citet{kothes01a} with the SNR.
A detailed study on the CO gas and its association with the \hii\ region
and the SNR will be presented in a separate paper (Jeong et al. in preparation).

%---------------------------------fig 6------------------------------------------------
\begin{figure}[t!]
\hspace{-0.2cm}
\includegraphics[scale=0.51]{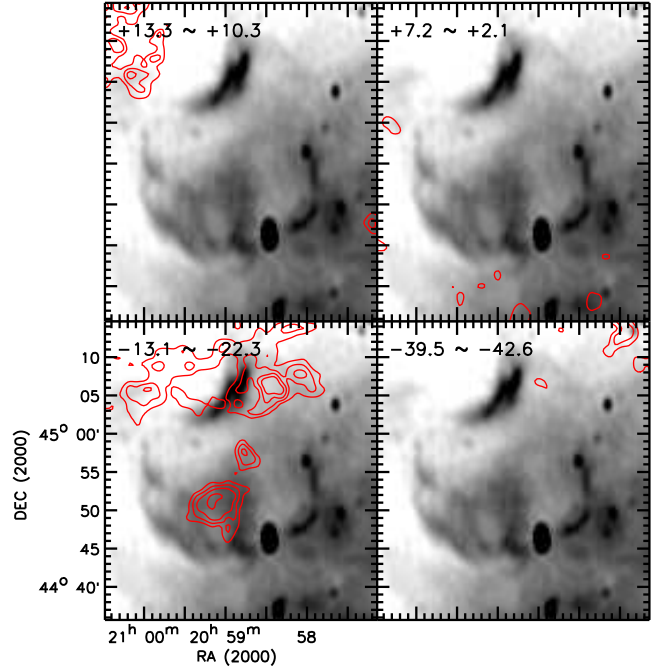}
\caption{$\rm{^{12}CO}~{\it J}$ = 1$-$0~average intensity maps of G85.9$-$0.6
(contour levels: 0.7, 1.1, 1.5, 2 K).
The gray-scale image is the CGPS 1420 MHz radio continuum image (scale range: 9.5$\sim$14 K).
\label{fig:g85_6}}
\end{figure}
%---------------------------------fig 6------------------------------------------------

\subsection{G85.9$-$0.6}
G85.9$-$0.6 is a shell-type SNR discovered recently by the CGPS.
\citeauthor{kothes01a} (\citeyear{kothes01a,kothes06}) suggested that the target might be
located in the low-density inter-arm region at $\sim$5 kpc
based on its faint radio and X-ray brightness.
There was no particular feature
observed from the \hi\ and polarized emission.
The distance based on the $\Sigma-D$ relation of
\cite{case98} is 19.6 kpc.

We detected CO molecular clouds at +12, +5, $-$20, and
$-$41 \kms~(Fig.~\ref{fig:aver_spec}). At $-$20 \kms,
there is a large molecular cloud near the northern
boundary of the SNR and small clouds in the central region
of the SNR (Fig.~\ref{fig:g85_6}).
However, there are no indications suggesting the interaction of
these clouds with the SNR.
The distribution of CO molecular clouds at other velocities
does not show any correlation with the radio morphology of the remnant.

%---------------------------------fig 7------------------------------------------------
\begin{figure}[t!]
\hspace{-0.2cm}
\includegraphics[scale=0.51]{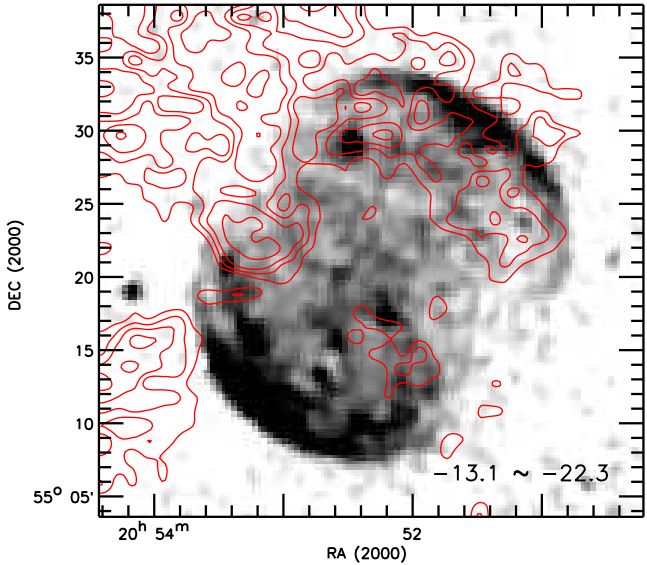}
\caption{$\rm{^{12}CO}~{\it J}$ = 1$-$0~average intensity maps of G93.3$+$6.9
(contour levels: 1.1, 1.3, 1.5, 1.8 K).
The gray-scale image is the WENSS 325 MHz radio continuum image (scale range: 9.5$\sim$14 Jy/beam) \citep{Renge90}.
\label{fig:g93}}
\end{figure}
%---------------------------------fig 7------------------------------------------------

\subsection{G93.3+6.9 (DA530)}
G93.3+6.9 is classified as bilateral
type with a well-defined shell morphology in the radio continuum \citep{gaensler98}.
This remnant is located at the northern edge of the
Cyg OB7 molecular cloud complex at 0.8 kpc that has
\vlsr\ = $-$7 to 0 \kms~(Huang \& Thaddeus 1998).
\cite{landecker99} found an \hi\ bubble at a distance of 3.5 kpc
that could have been produced by
stellar wind from the massive star progenitor.
\cite{foster03} estimated the distance of 2.2 kpc
by comparing their model of \hi\ column density distribution with the observed \hi\ data.
From Chandra X-ray observations,
\cite{jiang07} found an extended small hard X-ray
feature at the centre of the remnant and proposed that it would be
pulsar wind nebula.
OH maser observation was made toward the remnant with negative results
\citep{frail96}.

We detected CO emission at $-$6 to $+$5 \kms\ (Fig.~\ref{fig:aver_spec}).
Diffuse emissions at $\sim$$-$1.3 \kms~ appear to surround the
north-west of the remnant although the morphological correlation between
the CO and radio distribution is not clear (Fig.~\ref{fig:g93}).

\subsection{G94.0+1.0 (3C434.1)}
G94.0+1.0 is a shell type SNR with an asymmetric shape. In radio continuum emission,
the western area is faint and has no distinct shell-like feature in visible region,
whereas a bright semi-circular shell is evident in the east.
In CGPS 1420 MHz radio emission, there was no polarized emission
\citep{willis73,landecker85,kothes06}.
An \hi\ observation suggested that a
\hi\ stellar wind bubble at 4.5 kpc is surrounding the remnant
and that the progenitor could be an O4 star \citep{foster05}.
By using ROSAT data, he showed that bright, thermal
hard X-ray emission was detected toward the radio bright region
with good correlation.
An OH maser observation was made toward the remnant
with negative results \citep{frail96}.

\begin{figure}[t!]
\hspace{-0.2cm}
\includegraphics[scale=0.51]{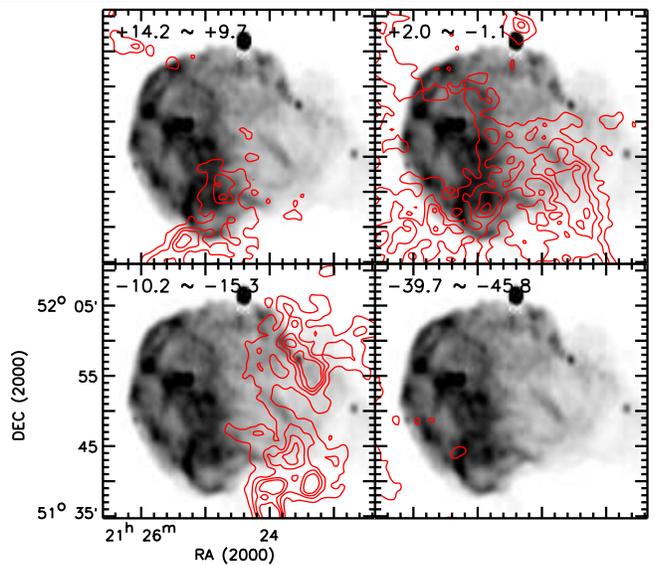}
\caption{$\rm{^{12}CO}~{\it J}$ = 1$-$0~average intensity maps of G94.0$+$1.0
(contour levels: 0.8, 1.5, 2, 2.5 K).
The gray-scale image is the CGPS 1420 MHz radio continuum image (scale range: 8$\sim$14 K).
\label{fig:g94}}
\end{figure}

We detected molecular gas over a wide velocity range from
$-70$ to $-40$ \kms~and from $-20$ to +20 \kms~(Fig.~\ref{fig:aver_spec}).
The positive velocity components (\vlsr~$\ge$ 0 \kms) and
large negative velocity components (\vlsr~$\le$ $-40$ \kms) do not show any correlated feature
between the MCs and the radio continuum emission.
We note that there is clear anti-correlation between the CO and radio distribution
at $\sim$$-$13 \kms~(Fig.~\ref{fig:g94}).
The large CO cloud is located in the western area of the weak radio continuum region
with well-defined spatial correlation.
Such anti-correlation is considerably similar to the case of
the SNR G109.1$-$1.0 (CTB109), for which
the interaction of the SNR with the molecular cloud occurs
\citep{tatematsu87,tatematsu90,sasaki06}.
A detailed study on the interaction between the molecular cloud and the SNR will be
reported in a separate paper \citep{jeong12}.
If the molecular cloud at $-$13 \kms~were associated with the SNR,
the distance to the SNR would be about 3.0 kpc.

%---------------------------------fig 8------------------------------------------------
%---------------------------------fig 8------------------------------------------------

%---------------------------------fig 9------------------------------------------------
\begin{figure}[t!]
\hspace{-0.2cm}
\includegraphics[scale=0.51]{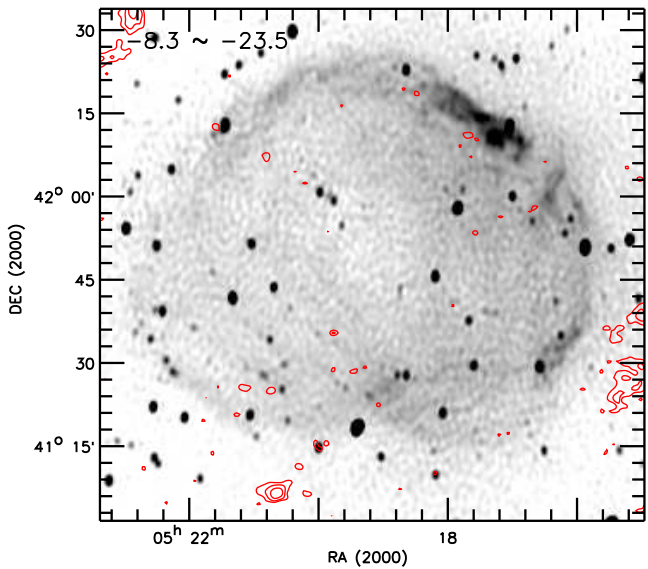}
\caption{$\rm{^{12}CO}~{\it J}$ = 1$-$0~average intensity maps of G166.2$+$2.5
(contour levels: 0.9, 1.3, 1.8 K).
The gray-scale image is the CGPS 1420 MHz radio continuum image (scale range: 4.6$\sim$6 K).
\label{fig:g166}}
\end{figure}
%---------------------------------fig 9------------------------------------------------

\subsection{G166.2+2.5 (OA184)}
G166.2+2.5 is a large radio source with low radio surface brightness,
and it was known as an evolved SNR \citep{willis73}.
However, this source has been recently classified as an
\hii\ region based on its flat radio spectral index,
missing X-ray emission, and non-significant polarization \citep{foster06}.
\cite{Huang86} noted that OA 184 is located
$\sim$2$^\circ$ west of VRO 45.05.01 and
suggested that both of them are associated with
a molecular cloud complex at \vlsr~= $-22$ \kms.
The cloud complex is located between the two sources and its velocity
is coincident with the velocities of the $H\alpha$~emission lines detected
toward them \citep{lozinskaya81}.
\cite{routledge86} suggested that an \hi~feature
at a \vlsr\ of $-$30 \kms~in the southwest is probably associated with the source.
OH maser observation was made toward the remnant with a negative result
\citep{frail96}.

We detected a few small ($\leq 5'$) CO clumps
from $-$26 to $-$9~\kms~(Fig.~\ref{fig:aver_spec}).
These clumps are located $\sim$10$'$ away from the radio boundary of the source
in the southeast and southwest (Fig.~\ref{fig:g166}).
They do not show any correlated features with the remnant.

\subsection{G179.0+2.6}
G179.0+2.6 is a faint SNR with a diameter of about 70$'$.
Three bright radio spots near the center of the remnant are radio
galaxies along the line of sight (\citeauthor{fuerst86} \citeyear{fuerst86,fuerst89}).

\begin{figure}[t!]
\hspace{-0.2cm}
\includegraphics[scale=0.51]{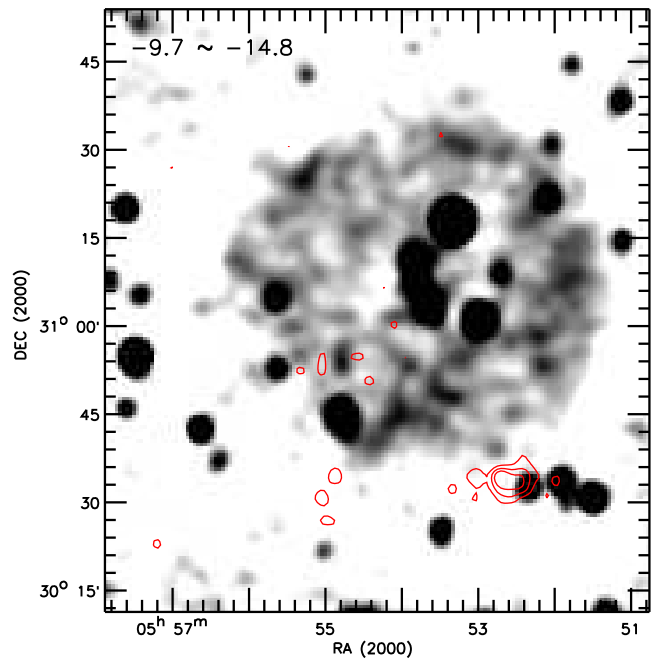}
\caption{$\rm{^{12}CO}~{\it J}$ = 1$-$0~average intensity maps of G179.0$+$2.6
(contour levels: 0.4, 0.8, 1.2 K).
The gray-scale image is the WENSS 325 MHz radio continuum image (scale range: 0.5$\sim$8 mJy/beam).
\label{fig:g179}}
\end{figure}

CO emission toward this remnant is extremely weak, as shown
in Fig.~\ref{fig:aver_spec}.
Only one molecular cloud with a 5$'$ diameter is detected
at $\sim$$-$12 \kms~(Fig.~\ref{fig:g179}).
This cloud lies near the southwest radio boundary; however,
no signature of the SNR-MC interaction is observed.

%---------------------------------fig 10------------------------------------------------
%---------------------------------fig 10------------------------------------------------

%---------------------------------fig 11------------------------------------------------
\begin{figure}[t!]
\hspace{-0.2cm}
\includegraphics[scale=0.51]{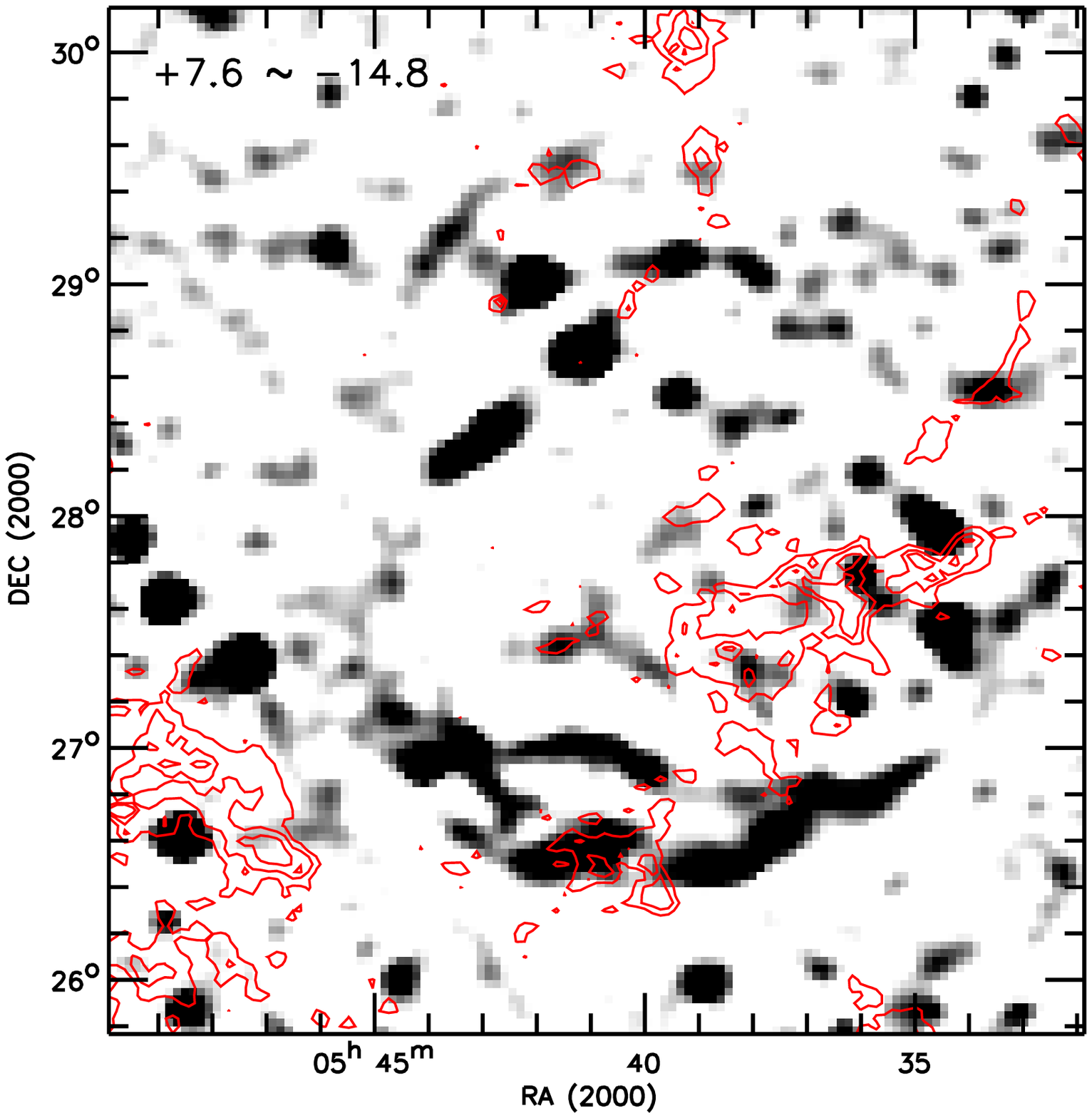}
\caption{$\rm{^{12}CO}~{\it J}$ = 1$-$0~average intensity maps of G180.0$-$1.7
(contour levels: 0.4, 0.8, 1.2 K).
The gray-scale image is the GB6 4850 MHz radio continuum image (scale range: 1.7$\sim$6.3 mJy/beam) \citep{condon94}.
\label{fig:g180}}
\end{figure}
%---------------------------------fig 11------------------------------------------------

%---------------------------------------------table start----------------------------------
%---------------------------------------------start table.Summary of Results ---------------------------------
\begin{table*}[t]
\small
\caption{Summary of Results\label{tbl-3}}
%\begin{center}
\begin{tabular}{@{}cccl@{}}
\tableline
Supernova   & ~~~ CO ~~~               & ~~ Kinematic ~~       &      \\
Remnant     & ~~~ velocity$^{a}$ ~~~   & ~~ distance$^{b}$ ~~  & ~~~~~~~~~~~~~~~~~~~~~~~~~~~~~~~~~~~~~~~~~~Note \\
            & ~~~ (km/s)~~~           & ~~ (kpc) ~~           &      \\
\tableline

G73.9 $+$0.9 & $+$3     &$\cdots$           & Extended, filamentary MC across the northern part of the remnant \\
             &          &                   & along the east-west direction.\\

G76.9 $+$1.0 & $\cdots$ & $\cdots$          & Large cloud across the SNR from southeast to northwest, but \\
             &          &                   & no obvious spatially-correlated features. \\

G84.2 $-$0.8 & $-$17    & 4.9              & Semi-circular CO cloud coincident with the western SNR \\
             &          &                  & boundary with enhanced radio brightness. \\

G85.4 $+$0.7 & $-$41    & 7.2              & Long, filamentary MC along the southwestern boundary of the SNR,  \\
             &          &                  & connected to the compact {\sc{Hii}} region G84.9+0.5 in the south.\\

G85.9 $-$0.6 & $\cdots$ & $\cdots$         & Several MCs in the field, but no obvious spatially-correlated features.   \\

G93.3 $+$6.9 & $-$2     & $\cdots$         & Large, diffuse MC surrounding the northeast boundary of the SNR.  \\

G94.0 $+$1.0 & $-$13    & 3.0              & Large molecular clouds blocking the western part of the SNR\\
             &          &                  & where the radio emission is faint.\\

G166.2 $+$2.5& $\cdots$ & $\cdots$         & Several small molecular clumps outside the SNR boundary. \\

G179.0 $+$2.6& $\cdots$ & $\cdots$         & Small molecular clumps in the southern part of the SNR, \\
             &          &                  & but no obvious spatially-correlated features. \\

G180.0 $-$1.7& $\cdots$ & $\cdots$         & Small MCs in the central area of the SNR without any correlation.     \\

G182.4 $+$4.3& $+$4     & $\cdots$         & Large, diffuse MC just outside the northwestern boundary of the SNR.   \\
\tableline
\end{tabular}
%\end{center}
\tablenotetext{a}{Velocity of the CO emission that is spatially-correlated with the SNR. }
\tablenotetext{b}{Kinematic distances corresponding to the CO velocity assuming the rotation curve of Brand \& Blitz (1993). For small CO velocities, no kinematic distances have been derived.}
\end{table*}
%---------------------------------------------end table.Summary of Results --------------------------------- 
%---------------------------------------------table end----------------------------------

%---------------------------------fig 12------------------------------------------------
\begin{figure}[t!]
\hspace{-0.2cm}
\includegraphics[scale=0.51]{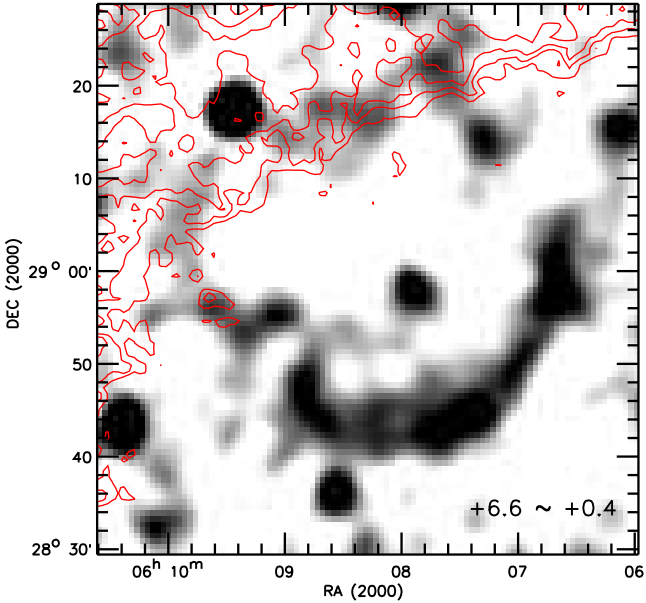}
\caption{$\rm{^{12}CO}~{\it J}$ = 1$-$0~average intensity maps of G182.4$+$4.3
(contour levels: 0.5, 0.8, 1.1, 1.5 K).
The gray-scale image is the GB6 4850 MHz radio continuum image (scale range: 0.38$\sim$9 mJy/beam).
\label{fig:g182}}
\end{figure}
%---------------------------------fig 12------------------------------------------------

\subsection{G180.0$-$1.7 (S147)}
G180.0$-$1.7 has large angular size of $\sim 3^\circ$ and remarkable optical
filaments \citep{van73}. \cite{anderson96} discovered a
pulsar 40$'$ away from the center to the west.
The age of the pulsar is estimated to be $3\pm0.4 \times 10^4$ yr \citep{kramer03}.
No OH maser emission was detected toward the remnant \citep{frail96}.

We found some molecular clouds at velocities
from $\sim$$-$14 to $+$5 \kms~ (Fig.~\ref{fig:aver_spec}).
Fig.~\ref{fig:g180} shows the average intensity map from
$-$14.8 to +7.6~\kms.
The cloud at $\sim$$-$11 \kms~is located $15'$ away toward the southeast
from the radio boundary.
CO emission at $\sim$$-$5 \kms~is detected near the radio filaments
in the south and shows filamentary and clumpy morphology.
The cloud at $\sim +1~$\kms~ is located in the western part of
the remnant where radio brightness is very faint.
Although we detected molecular clouds near the SNR,
none of them show correlated features with the remnant.

\subsection{G182.4+4.3}
G182.4+4.3 is a shell-type SNR with a highly polarized,
bright partial shell in the southwest \citep{kothes98}.
The southern shell is bright and circular, whereas
the northern shell is faint and flattened.
No X-ray emission was detected toward the SNR.

We detected CO emissions at velocities
from $-$9 to $+$10 \kms~(Fig.~\ref{fig:aver_spec}).
Most CO emissions arise outside the north and northeast shell of the remnant
where the radio brightness is significantly fainter than that of the south shell.
The boundary of the molecular cloud at $\sim$$+$4 \kms~
matches well with the radio boundary of the remnant (Fig.~\ref{fig:g182}).
It is possible that the molecular cloud is blocking the remnant,
although no direct evidence for their interaction has been detected.

%---------------------------------------------Result end----------------------------------------

%-------------------------------------------Summary start-----------------------------
\section{Summary}\label{s:summary}
We carried out $\rm{^{12}CO}~{\it J}$ = 1$-$0~line survey of SNRs
using the SRAO 6-m telescope. We observed
11 SNRs between $l$ = 70$^\circ$ and 190$^\circ$:
G73.9$+$0.9, G76.9$+$1.0, G84.2$-$0.8, G85.4$+$0.7, G85.9$-$0.6, G93.3$+$6.9 (DA530),
94.0$+$1.0 (3C434.1), 166.2$+$2.5 (OA184), 179.0$+$2.6, 180.0$-$1.7 (S147), and G182.4$+$4.3.
The mapping is performed
either in half-beam or full-beam sampling
to cover the full extents of individual SNRs in the radio continuum.
The total number of spectra is about 45,000.
We summarize our main results below:\\

1. We found CO molecular clouds
having spatial correlations with radio emission in six SNRs:
G73.9$+$0.9, G84.2$-$0.8, G85.4$+$0.7, G93.3$+$6.9, G94.0$+$1.0, and G182.4$+$4.3.
However, no strong evidence for the interaction, e.g., broad wings or high-velocity
shocked gas, was detected for these objects.
We did not detect molecular clouds having
any obvious spatial correlation with radio emission in the other SNRs.
Table~\ref{tbl-3} summarizes the results in which the second column lists the
velocities of the possibly associated CO features and their kinematic distances.
The fourth column gives a short summary on individual SNRs. \\

2. Two SNRs are particularly interesting.
In G85.4$+$0.7, there is a well-defined filamentary molecular cloud
that matches well with the southern
boundary of the SNR at $-41$ \kms.
Another interesting SNR is 3C434.1. It has molecular clouds
in the western part of the SNR, where
the radio emission is weak. Such anti-correlation between a molecular cloud and
radio emission was previously found in the SNR CTB109 (G109.1$-$1.0),
and the their association is likely.
Further observations are needed to confirm the interaction between
molecular clouds and these SNRs.
%-------------------------------------------Summary end-----------------------------

%----------------------------------------start: acknowledgements -----------------------
\acknowledgments
This work is supported in part by the Korean Research
Foundation under grant KRF-2008-313-C00372.
This research used the facilities of the Canadian Astronomy Data Centre operated
by the National Research Council of Canada with the support of the Canadian Space Agency.
The GB6 4850 MHz and WENSS 325 MHz radio images were
retrieved from NASA's \emph{SkyView} and we acknowledge the use of NASA's SkyView facility
(http://skyview.gsfc.nasa.gov) located at NASA Goddard Space Flight Center.
%----------------------------------------end: acknowledgements -----------------------

%\bibliographystyle{spr-mp-nameyear-cnd}
%\bibliography{myref}
%\bibliography{biblio-u1}

\begin{thebibliography}{}
\bibitem[\protect\citeauthoryear{Abdo et al.}{2009}]{abdo09} Abdo,~A.A. et al.: \apjl\ \textbf{706}, L1 (2009)
\bibitem[\protect\citeauthoryear{Acciari et al.}{2009}]{acciari09} Acciari,~V.A. et al.: \apjl\ \textbf{698}, L133 (2009)
\bibitem[\protect\citeauthoryear{Aharonian et al.}{2008}]{aharonian08} Aharonian,~F. et al.: \aap\ \textbf{490}, 685 (2008)
\bibitem[\protect\citeauthoryear{Anderson et al.}{1996}]{anderson96} Anderson,~S.B., Cadwell,~B.J., Jacoby,~B.A., Wolszczan,~A., Foster,~R.S., \& Kramer,~M.: \apj\ \textbf{468}, L55 (1996)
\bibitem[\protect\citeauthoryear{Brand \& Blitz}{1993}]{brand93} Brand, J., \& Blitz, L.: \aap\ \textbf{275}, 67 (1993)
\bibitem[\protect\citeauthoryear{Byun}{2004}]{byun04} Byun, D.-Y. ``Development of the SRAO 6-meter Telescope Control System and Molecular Line Observation of Supernova remnants" PhD thesis, Seoul National University (2004)
\bibitem[\protect\citeauthoryear{Byun et al.}{2006}]{byun06} Byun, D.-Y., Koo, B.-C., Tatematsu, K. \& Sunada, K.: \apj\ \textbf{637}, 283 (2006)
\bibitem[\protect\citeauthoryear{Case et al.}{1998}]{case98} Case, G.L., \& Bhattacharya, D.: \apj\ \textbf{504}, 761 (1998)
\bibitem[\protect\citeauthoryear{Chen et al.}{2004}]{chen04} Chen, Y., Su, Y., Slane, P., \& Wang, Q. D.: \apj\ \textbf{616}, 885 (2004)
\bibitem[\protect\citeauthoryear{Choi et al.}{2003}]{choi03} Choi, H.-K., Byun, D.-Y., \& Koo, B.-C.: International Journal of Infrared and Millimeter Waves, \textbf{24}, 683 (2003)
\bibitem[\protect\citeauthoryear{Claussen et al.}{1997}]{claussen97} Claussen, M.J., Frail, D.A., Goss, W.M., \& Gaume, R.A.: \apj\ \textbf{489}, 143 (1997)
\bibitem[\protect\citeauthoryear{Condon et al.}{1994}]{condon94} Condon, J.J., Broderick, J.J., Seielstad, G.A., Douglas, K. \& Gregory, P.C.: \aj\ \textbf{107},1829 (1994)
\bibitem[\protect\citeauthoryear{DeNoyer}{1979}]{denoyer79} DeNoyer, L.K.: \apjl\ \textbf{228}, L41 (1979)
\bibitem[\protect\citeauthoryear{Elitzur}{1976}]{elitzur76} Elitzur, M.: \apj\ \textbf{203}, 124 (1976)
\bibitem[\protect\citeauthoryear{Feldt \& Green}{1993}]{feldt93} Feldt, C., \& Green, D.A.: \aap\ \textbf{274}, 421 (1993)
\bibitem[\protect\citeauthoryear{Foster}{2005}]{foster05} Foster, T.: \aap\ \textbf{441}, 1043 (2005)
\bibitem[\protect\citeauthoryear{Foster et al.}{2007}]{foster07} Foster, T.J., Kothes, R., Kerton, C.R., \& Arvidsson, K.: \apj\ \textbf{667}, 248 (2007)
\bibitem[\protect\citeauthoryear{Foster et al.}{2006}]{foster06} Foster, T., Kothes, R., Sun, X.H., Reich, W. \& Han, J.L.: \aap\ \textbf{454}, 517 (2006)
%\bibitem[\protect\citeauthoryear{Foster et al.}{2006b}]{foster06b} Foster, T. \& MacWilliams, J.: \apj\ \textbf{644}, 214 (2006b)
\bibitem[\protect\citeauthoryear{Foster et al.}{2003}]{foster03} Foster, T., \& Routledge, D.: \apj\ \textbf{598}, 1005 (2003)
\bibitem[\protect\citeauthoryear{Frail et al.}{1996}]{frail96} Frail, D.A., Goss, W.M., Reynoso, E.M., Ciacani, E.B., Green, A.J., \& Otrupcek, R.: \aj\ \textbf{111}, 1651 (1996)
\bibitem[\protect\citeauthoryear{Fuerst et al.}{1986}]{fuerst86} Fuerst, E., \& Reich, W.: \aap\ \textbf{154}, 303 (1986)
\bibitem[\protect\citeauthoryear{Fuerst et al.}{1989}]{fuerst89} Fuerst, E., Reich, W., Kuhr, H., \& Stickel, M.: \aap\ \textbf{223}, 66 (1989)
\bibitem[\protect\citeauthoryear{Gaensler}{1998}]{gaensler98} Gaensler, B.M.: \apj\ \textbf{493}, 781 (1998)
\bibitem[\protect\citeauthoryear{Green}{2004}]{green04} Green, D.A.: Bulletin of the Astronomical Society of India \textbf{32}, 335 (2004)
\bibitem[\protect\citeauthoryear{Green}{2009}]{green09} Green, D.A.: Bulletin of the Astronomical Society of India \textbf{37}, 45 (2009)
\bibitem[\protect\citeauthoryear{Higgs et al.}{1983}]{higgs83} Higgs, L.A., Landecker, T.L. \& Roger, R.S.: \aj\ \textbf{88}, 97 (1983)
\bibitem[\protect\citeauthoryear{Huang \& Thaddeus}{1986}]{Huang86} Huang, Y.-L., \& Thaddeus, P.: \apj\ \textbf{309}, 804 (1986)
\bibitem[\protect\citeauthoryear{Jackson et al.}{2006}]{Jackson06} Jackson, J.M. et al.: \apjs\ \textbf{163}, 145 (2006)
\bibitem[\protect\citeauthoryear{Jeong et al.}{2012}]{jeong12} Jeong, I.-G., Koo, B.-C., Cho, W.-K., Kramer, C., Stutzki, J. \& Byun, D.-Y.: in preparation (2012)
\bibitem[\protect\citeauthoryear{Jiang et al.}{2007}]{jiang07} Jiang, B., Chen, Y., \& Wang, Q.D.: \apj\ \textbf{670}, 1142 (2007)
\bibitem[\protect\citeauthoryear{Jiang et al.}{2010}]{jiang10} Jiang, B., Chen, Y., Wang, J., Su, Y., Zhou, X., Safi-Harb, S., \& DeLaney, T.: \apj\ \textbf{712}, 1147 (2010)
\bibitem[\protect\citeauthoryear{Koo}{2003}]{koo03a} Koo, B.-C.: Shocked Atomic and Molecular Gas in Supernova Remnants (ASP Conf. Ser. \textbf{289}), ed. S. Ikeuchi, J. Hearnshaw, \& T. Hanawa (San Francisco: ASP), 199 (2003)
\bibitem[\protect\citeauthoryear{Koo et al.}{2003}]{koo03b} Koo, B.-C., Park, Y.-S., Hong, S.-S., Yun, H.-S., Lee, S.-G., Byun, D.-Y., Lee, J.-W., Choi, H.-K., Lee, S.-S., Yoon, Y.-Z., Kim, K.-T., Kang, H.-W., \& Lee, J.-E.: JKAS~\textbf{36}, 43 (2003)
\bibitem[\protect\citeauthoryear{Kothes et al.}{2006}]{kothes06} Kothes, R., Fedotov, K., Foster, T.J. \& Uyaniker, B.: \aap\ \textbf{457}, 1081 (2006)
\bibitem[\protect\citeauthoryear{Kothes et al.}{1998}]{kothes98} Kothes, R., Fuerst, E., \& Reich, W.: \aap\ \textbf{331}, 661 (1998)
\bibitem[\protect\citeauthoryear{Kothes et al.}{2001a}]{kothes01a} Kothes, R., Landecker, T.L., Foster, T., \& Leahy, D.A.: \aap\ \textbf{376}, 641 (2001a)
\bibitem[\protect\citeauthoryear{Kothes et al.}{2003}]{kothes03} Kothes, R., Reich, W., Foster, T. \& Byun, D.-Y.: \apj\ \textbf{588}, 852 (2003)
\bibitem[\protect\citeauthoryear{Kothes et al.}{2001b}]{kothes01b} Kothes, R., Uyaniker, B. \& Pineault, S.: \apj\ \textbf{560}, 236 (2001b)
\bibitem[\protect\citeauthoryear{Kramer et al.}{2003}]{kramer03} Kramer, M., Lyne, A.G., Hobbs, G., Lohmer, O., Carr, P., Jordan, C., \& Wolszczan, A.: \apjl\ \textbf{593}, L31 (2003)
\bibitem[\protect\citeauthoryear{Landecker et al.}{1985}]{landecker85} Landecker, T.L., Higgs, L.A., Roger, R.S.: \aj\ \textbf{90}, 1082 (1985)
\bibitem[\protect\citeauthoryear{Landecker et al.}{1993}]{landecker93} Landecker, T.L., Higgs, L.A. \& Wendker, H.J.: \aap\ \textbf{276}, 522 (1993)
\bibitem[\protect\citeauthoryear{Landecker et al.}{1999}]{landecker99} Landecker, T.L., Routledge, D., Reynolds, S.P., Smegal, R.J., Borkowski, K.J., \& Seward, F.D.: \apj\ \textbf{527}, 866 (1999)
\bibitem[\protect\citeauthoryear{Lee et al.}{2002}]{lee02} Lee, J.-W., Han, S.-T., Byun, D.-Y., Koo, B.-C., \& Park, Y.-S.: International Journal of Infrared and Millimeter Waves \textbf{23}, 47 (2002)
\bibitem[\protect\citeauthoryear{Lee et al.}{2012}]{lee12} Lee, J.-J., Koo, B.-C., 
Snell, R. L., Yun, M. S., Heyer, M. H., \& Burton, M. G.: \apj\ \textbf{749}, 34 (2012)
\bibitem[\protect\citeauthoryear{Lee et al.}{2008}]{lee08} Lee, J.-J., Koo, B.-C., Yun, M. S., Stanimirovi\'c, S., Heiles, C., \& Heyer, M. H.: \aj\ \textbf{135}, 796 (2008)
\bibitem[\protect\citeauthoryear{Lockett et al.}{1999}]{lockett99} Lockett, P., Gauthier, E., \& Elitzur, M.: \apj\ \textbf{511}, 235 (1999)
\bibitem[\protect\citeauthoryear{Lozinskaya}{1981}]{lozinskaya81} Lozinskaya, T.A.: SvAL \textbf{7}, 17 (1981)
\bibitem[\protect\citeauthoryear{Matthews et al.}{1980}]{matthews80} Matthews, H.E. \& Shaver, P.A.: \aap\ \textbf{87}, 255 (1980)
\bibitem[\protect\citeauthoryear{Pineault et al.}{1996}]{pineault96} Pineault, S., Gaumont-Guay, S., \& Madore, B.: \aj\ \textbf{112}, 201 (1996)
\bibitem[\protect\citeauthoryear{Reich et al.}{1986}]{reich86} Reich, W., Fuerst, E., Reich, P., Sofue, Y., \& Handa, T.: \aap\ \textbf{155}, 185 (1986)
\bibitem[\protect\citeauthoryear{Reach et al.}{1999}]{reach99} Reach, W.T. \& Rho, J.: \apj\ \textbf{511}, 836 (1999)
\bibitem[\protect\citeauthoryear{Rengelink et al.}{1990}]{Renge90} Rengelink, R.B., Tang, Y., de Bruyn, A.G., Miley, G.K., Bremer, M.N., Roettgering, H.J.A. \& Bremer, M.A.R.: \aaps\ \textbf{124}, 259 (1997)
\bibitem[\protect\citeauthoryear{Routledge et al.}{1986}]{routledge86} Routledge, D., Landecker, T.L., \& Vaneldik, J.F.: \mnras\ \textbf{221}, 809 (1986)
\bibitem[\protect\citeauthoryear{Sasaki et al.}{2006}]{sasaki06} Sasaki, M., Kothes, R., Plucinsky, P.P., Gaetz, T.J. \& Brunt, C.M.: \apjl\ \textbf{642}, L149 (2006)
\bibitem[\protect\citeauthoryear{Snell et al.}{2005}]{snell05} Snell, R.L., Hollenbach, D., Howe, J.E., Neufeld, D.A., Kaufman, M.J., Melnick, G.J., Bergin, E.A., \& Wang, Z.: \apj\ \textbf{620}, 758 (2005)
\bibitem[\protect\citeauthoryear{Tatematsu et al.}{1990}]{tatematsu90} Tatematsu, K., Fukui, Y., Iwata, T., Seward, F.- D., \& Nakano, M.: \apj\ \textbf{351}, 157 (1990)
\bibitem[\protect\citeauthoryear{Tatematsu et al.}{1987}]{tatematsu87} Tatematsu, K., Fukui, Y., Nakano, M., Kogure, T., Ogawa, H. \& Kawabata, K.: \aap\ \textbf{184}, 279 (1987)
\bibitem[\protect\citeauthoryear{Taylor et al.}{2003}]{taylor03} Taylor, A.R., et al.: \aj\ \textbf{125}, 3145 (2003)
\bibitem[\protect\citeauthoryear{van den Bergh et al.}{1973}]{van73} van den Bergh, S., Marscher, A.P., \& Terzian, Y.: \apjs\ \textbf{26}, 19 (1973)
\bibitem[\protect\citeauthoryear{Wardle}{1999}]{wardle99} Wardle, M.: \apj\ \textbf{525}, L101 (1999)
\bibitem[\protect\citeauthoryear{Wardle, M., \& Yusef-Zadeh}{2002}]{wardle02} Wardle, M., \& Yusef-Zadeh, F.: Science \textbf{296}, 2350 (2002)
\bibitem[\protect\citeauthoryear{Willis}{1973}]{willis73} Willis, A.G.: \aap\ \textbf{26}, 237 (1973)
%\bibitem[\protect\citeauthoryear{Wilner et al.}{1998}]{wilner98} Wilner, D. J., Reynolds, S. P., \& Moffett, D. A.: \aj\ \textbf{115}, 247 (1998)
\end{thebibliography}

%----------------------------------------thebibliography start-----------------------

%-----------------------------------thebibliography end -------------------------------------

\end{document}